\documentclass[showpacs,floatfix,aps,prl,twocolumn,superscriptaddress]{revtex4}
\usepackage{latexsym}
\usepackage{times}
\usepackage{graphicx}

\begin{document} 

\author{M. Cosentino Lagomarsino}
\affiliation{UMR 168 / Institut Curie, 26 rue d'Ulm 75005 Paris, France}
\email[ e-mail address: ]{mcl@curie.fr}
\author{P. Jona} 
\affiliation{Politecnico di Milano, Dip. Fisica, Pza Leonardo Da Vinci
  32, 20133 Milano, Italy} 
%
\author{B.  Bassetti} 
\affiliation{Universit\`a degli Studi di Milano, Dip.
    Fisica, and I.N.F.N. Via Celoria 16, 20133 Milano, Italy } 
\email[e-mail address: ]{ bassetti@mi.infn.it }


\date{\today}

\title{The Logic Backbone of a Transcription Network}

\begin{abstract}
  A great part of the effort in the study of coarse grained models of
  transcription networks concentrates on their dynamical features. In this
  letter, we consider their \emph{equilibrium} properties, showing that the
  backbone underlying the dynamic descriptions is an optimization problem.  It
  involves $N$ variables, the gene expression levels, and $M$ constraints, the
  effects of transcriptional regulation.  In the case of Boolean variables and
  constraints, we investigate the structure of the solutions, and derive phase
  diagrams.  Notably, the model exhibits a connectivity transition between a
  regime of simple gene control, where the input genes control O(1) other
  genes, to a regime of complex control, where some ``core'' input genes
  control O$(N)$ others.
\end{abstract}

\maketitle

\begin{figure}[htbp]
  \centering \includegraphics[width=.35\textwidth]{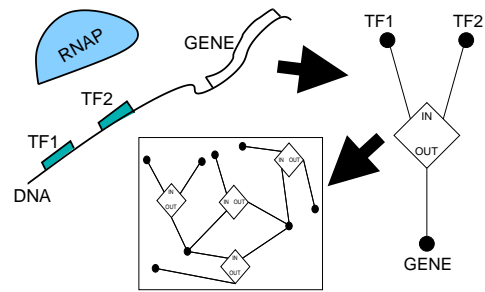}
  \caption{Schematics of the representation of a transcription
    network. Each signal integration function at the \emph{cis-}regulatory
    region of a gene corresponds to a constraint on the gene expression
    variables.  Bottom: example of factor graph of GR1 for $ K_b = K = 2$.
    Each diamond node represents a constraint, while each black circle is a
    variable.}
  \label{fig:Fagraph}
\end{figure}

\paragraph{Introduction.}
Identity, response and architecture of a living system are central topics of
molecular biology. Presently, they are largely seen as a result of the
interplay between a gene repertoire and the regulatory
machinery~\cite{BLA+04,HCP04}.  Gene transcription in mRNA form is an
important step in this process.
At this level, the regulatory machinery is embodied by the transcription
factors, proteins that bind to special sites along DNA and control the
activity of RNA polymerase~\cite{Pta92,BGH03} (Fig.~\ref{fig:Fagraph}).
This process is referred to as signal integration.  
Together, the \emph{cis}-regulatory regions establish a set of
interdependencies between transcription factors and genes, including other
transcription factors: a ``transcription network''~\cite{BLA+04}. Some of such
networks of living organisms are now being explored
experimentally~\cite{HCP04}, and show a modularity that has important
biological implications~\cite{SMM+02}.
Understanding the gene expression patterns determined by these networks is an
enormous challenge. The problem is that transcription networks are fairly
large, so that a coarse graining is needed. This fact has many consequences,
mainly related to the dynamics.
For example, the pioneering approach of Kauffman~\cite{Kau93} suggesting a
synchronous deterministic update for a Boolean (i.e. on/off) representation of
the network is still being debated~\cite{Ger04}.
Microscopically, it is well accepted that the Gillespie
algorithm~\cite{GD77,MA97} correctly describes the events of chemical kinetics
involved. On the other hand, with a mesoscopic average in time, it is not
clear what the emergent time scales might be.

\begin{figure}[htbp]
  \includegraphics[width=.35\textwidth]{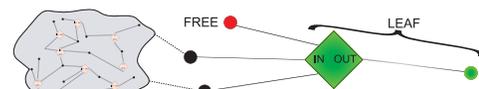}
  \caption{Example of a leaf. The free gene that regulates it will
    not be in the core. The other two transcription factors are connected to
    the rest of the network, represented by a cartoon.}
  \label{fig:foglia}
\end{figure}

We approach this problem with a model based on two features. Firstly, it
focuses, rather than on dynamics, on the compatibility between gene expression
patterns and signal integration functions. Simply put, a cell with $N$ genes
can express them in exponentially many ways, $2^N$ in the Boolean
representation. However, the cell never explores all the patterns of
expression. It only knows clusters of correlated configurations.  An
elementary example is the cI-cro switch of $\lambda$-phage~\cite{Pta92}.
Looking at the system, one could observe the states 10, 01 or perhaps 00, but
not 11.  One can think that in larger systems many configurations are ruled
out for the same compatibility reasons. Secondly, the model takes explicitly
into account that some genes are essentially ``free'' from the point of view
of transcription (Fig.~\ref{fig:foglia}). This fact is evident looking at the
available data~\cite{SMM+02}. While the biological situation is more complex,
we regard these genes as input receptors, connected to external stimuli.  The
simplest formulation (GR1, from Gene-Regulation) assumes Boolean variables and
functions.  We use it to investigate theoretically the control exerted by the
free genes on the expression patterns for large $N$, and for random
realizations of the constraints~\cite{M02}.
%
Analysis of the satisfying configurations leads to the introduction of a
``core'' of network variables.
Depending on the number of free genes in the core and the connectivity of the
constraints, there are three distinct regimes of gene control.
In the first regime, the core is empty.  Each free gene controls the state of
a small number of genes (``simple control'' phase).  In the second regime
(``complex control''), the core contains free genes that control, both
directly and indirectly, order $N$ others. Thus, in the complex control phase,
the free core genes can be interpreted as the subset of genes that determine a
choice of an expression program.  In the third regime, there are no free genes
in the core, and the system cannot control the simultaneous expression of all
its genes. The transition can be tuned by varying the connectivity and the
number of constraints.

\paragraph{Model.}
The two main ingredients of our representation of a transcription network are:
(i) A set of $N$ discrete variables $\{ x_i \}_{i=1..N}$ associated to genes or
operons (identified with their transcripts and protein products). These
variables represent the expression levels.
(ii) A set of $M$ interactions, representing the signal integration, $\{
I_b(x_{b_0},x_{b_1},.., x_{b_{K_b}}) \}_{b=1..M}$, with $\gamma = M/N \le 1$.
It is useful to represent variables and interactions in a so-called factor
graph, as illustrated in Fig.~\ref{fig:Fagraph}. Note that the constraints
$I_b$ contain the topology of the graph.
%
The $x_i$ represent real or coarse-grained expression levels and can take
values in $\{0,..,q\}$ or even continuous ones. In general, using the Shea and
Ackers model of gene activation by recruitment~\cite{SA85,BGH03}, one can
construct a local free energy associated to each signal integration node, that
generates the constraints~\cite{CLB05}.
Here we consider the simplest possible scenario, GR1, treating the
expression levels as Boolean variables (i.e. setting $q=1$), and the signal
integration functions as Boolean functions $\{ f_b(x_{i(b,1)}, ..,
x_{i(b,K_b)}) \}_{b=1..M}$.  The coordinates $i(b,l)$ point at the
variable occupying place $l$ in the $b$th constraint. 
The expression
\begin{equation}
   x_{i(b,0)} = f_b(x_{i(b,1)}, .., x_{i(b,K_b)}) ,
   \label{eq:fp}
\end{equation}
imposes that the variable $ x_{i(b,0)}$  is the output of the function $f_b$. 
For example, let us consider a graph with three variables and one constraint,
labeled by $b=1$.  Supposing the first two variables regulate the third,
$K_1=2$, $i(1,0)=3;\, i(1,1)=1;\, i(1,2)=2$.  If, for instance, the
transcription can occur only when both regulators are present, then
$f_1(x_1,x_2) = 1$ only if $x_1,x_2 =1$ (Boolean AND function).
The local connectivity of a function node is $k_{b} = 1+ K_{b}$.  $K_{b}$ is
called the ``in-degree''. The factor graph is also associated to an
``out-degree'' $C_{i} \equiv c_{i} - 1 $, where $c_{i}$ is the number of
functions connected to $x_{i}$.
The fact that variables and constraints are Boolean make GR1 an optimization
problem of the satisfiability type (Sat)~\cite{M02}. A very special one, given
the structure of the signal integration functions.  The properties of the
model depend on the class of graphs and Boolean functions used. The results
presented in this work hold for a rather large class of Boolean functions (see
appendix A).

\paragraph{Structure of the solution space.}
The phase-space structure has never been explored for this particular case.
We set out to analyze the number of compatible configurations $\mathcal{N}$,
for large $N$ and $M$ and random instances of the problem.
To this end, consider the following argument, which focuses on the control
exerted by the $N-M$ ``free'' input genes on the compatible solutions.
Together with the free genes, the network contains some genes which are
regulated but do not regulate any other.  We can refer to them as ``leaves''
(Fig~\ref{fig:foglia}). Given a realization, a leaf can always be adjusted to
the output value of its function, which is then satisfied. Let us now remove
from the graph each leaf and iterate this procedure (a variation of the
so-called ``leaf removal'' algorithm~\cite{MRZ03}).  There are two possible
outcomes: either erasing all the graph, leaving the free genes as isolated
points, or stopping at a core of constraints that contains loops. In other
words, the algorithm identifies the tree-like components of the graph
connected to outputs.  The core will be composed of $N_C$ genes and $M_C$
constraints. Let us now imagine to have a compatible configuration, flip a
free gene, and try to construct another compatible configuration.  In the case
where the core is empty, since the graph is tree-like, it will always be
possible to perform this operation by local rearrangements, which propagate
the output of the functions. Thus, flipping all the free genes, we find
$2^{N-M}$ satisfying configurations. In the presence of a core, because of the
loops, flipping a free gene of the core will in general rearrange all the core
genes, and is not guaranteed to lead to a new satisfying state. In fact, it is
not granted there will be a compatible state to start with. Provided there is,
the output propagation procedure can be applied to the non-core free genes to
construct another.
%
Thus, in general the $N-M$ degrees of freedom given by the free genes cannot
guarantee a solution.  The relevant parameter is the number of core free genes
$\Delta_{C} = N_{C} - M_{C}$. Let us for the moment restrict to the case of
fixed in-degree (``$K$-GR1''). In appendix B, we show that the average of
$\mathcal{N}$ on the class of all Boolean functions is $2^{N-M}$. Thus: (a)
if the core is empty, the number of compatible configurations constructed by
flipping the free genes are on average all the possible ones.  (b) If the core
is not empty, in the average case it will still be possible to construct
$2^{N-M}$ solutions by flipping the free genes.  If $\Delta_{C} > 0$, there
will be $2^{\Delta_{C}}$ clusters of solutions, and (c) in the case where the
core contains no free genes there will be generally contradictions.  We can
thus distiguish the three regimes: (a) simple control, (b) complex
control, (c) no control.
Considering ensembles of random graphs, the regimes above depend on the value
of $N_C(N), M_C(M)$, so that a proper order parameter to adopt is $\gamma =
M/N$~\cite{M02}. The phase diagram can be explored studying the rank and the
kernel of the connectivity matrix in the ensemble.
%
\paragraph{Example: the case of Poisson-distributed $c_{i}$.} 
This is the simplest ensemble to consider, where $p(c) = \frac{(k
  \gamma)^{c}}{c!}  e^{-k \gamma}$~\cite{MRZ03}.
This probability distribution doesn't exclude that genes may appear
in the functions that regulate them, leaving some freedom of choice. 
In the simplest case, one finds $N_C (\gamma) = N(m - k \gamma (1 -m)m^{k-1})$
and $M_C (\gamma)= N( \gamma m^k)$, where $m$ is defined by the relation
$m(k)+ e^{-k \gamma m(k)^{k-1}}- 1 = 0$.
%
%
%
%
This gives the phase diagram as a function of $\gamma$.  For $\gamma<\gamma_d$
there is simple control. For $\gamma_d<\gamma<\gamma_c$ complex control, and
for $\gamma>\gamma_c$ no control. For example, for $4$-GR1, $\gamma_d \simeq
0.776$ and $\gamma_c \simeq 0.977$.  The regimes of gene control
correspond to thermodynamic phases,
commonly referred to as SAT, HARD-SAT, and UNSAT phase
respectively~\cite{MPZ02}.
\begin{figure}[htbp]
  \centering
  \includegraphics[width=.38\textwidth]{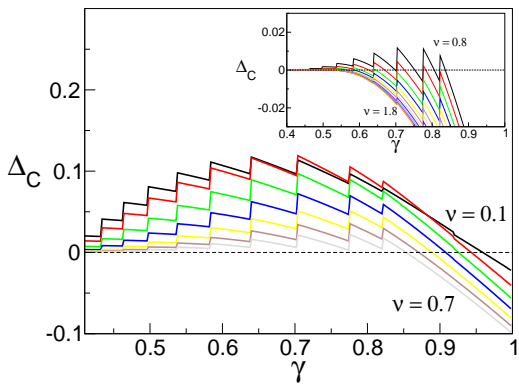}
  \caption{$\Delta_C$ as a function of $\gamma$ for different values of $\nu$
    in the multi-Poisson case. The discrete jumps are due to the onset of
    complex control phases for the different values of $k$. $\Delta_C$ can
    become negative many times, giving rise to reentrant no control phases
    (inset).  The figure refers to a connectivity distribution with a cutoff
    at $k = 12$.}
  \label{fig:mp-delta}
\end{figure}
Furthermore, it is possible to show rigorously the clustering of solutions
argued above. In the simple control phase, one cluster contains all the
solutions, and a free gene controls $O(1)$ other genes. The reason for this is
that, for Poisson distributed out-degree, the average number of controlled
genes is finite $(c\gamma)$, while the number of free genes is extensive.
Conversely, in the complex control phase, the free genes within the core
control $O(N)$ other genes (while there is still $O(1)$ control outside of the
core).
From a physical point of view, the clusters are separated by an extensive
distance, i.e. by free energy barriers.  The number of clusters is related to
the (computational) complexity $\Sigma$ of the system, defined by the relation
$ \log\mathcal{N} \sim N (\Sigma + S)$. Here $S$, the entropy, measures the
width of each cluster, while $\Sigma$ ``counts'' the number of clusters.
Therefore, by definition, $\Sigma$ is directly related to $\Delta_{C}$, i.e.
to the partitioning of the free genes in and out of the core.  How the system
explores (or not) these clusters depends on details of its dynamics.
Generically, one can say that the dynamics in a cluster will be residual: many
variables are fixed, the others can change.  This matches a qualitative
feature of many cells, where some genes are constantly expressed, and the rest
vary~\cite{nota_sav}.

\paragraph{Multi-Poisson phase diagram.}
While the fixed $k$ case is useful to get some theoretical insight, the known
transcription networks are far from having fixed in-degree. For example, in
E.~coli, the in-degree has Poisson distribution, while the out-degree
resembles a power law.
For this reason, a biologically more interesting case is when both the in- and
the out-degree vary along the network. Considering $p(k|c) =
\frac{(k\gamma)^c}{c!} e^{-(k\gamma) }$, the conditioned probability that a
variable is in $c$ clauses of the $k$ kind, we have $p(c) =
\sum_k\frac{(k\gamma)^c}{c!}  e^{-(k\gamma) } p(k)$. The leaf removal
equations can be applied separately to sets of clauses with a given
connectivity, defining $ N_C \equiv <N_C>_{k}$ and $M_C \equiv <M_C>_{k}$,
where $<X>_{k}= \sum_p p(k) X(k)$.
%
Choosing $p(k) = Z^{-1}(\nu) e^{-\nu k}$ does not affect the exponential
asymptotic decay of $p(c)$ for large $c$. We can call this case multi-Poisson,
as the graph is a superposition, following a Poisson distribution, of graphs
with fixed in-degree and Poisson-distrubuted out-degree.
\begin{figure}[htbp]
  \centering
   \includegraphics[width=.37\textwidth]{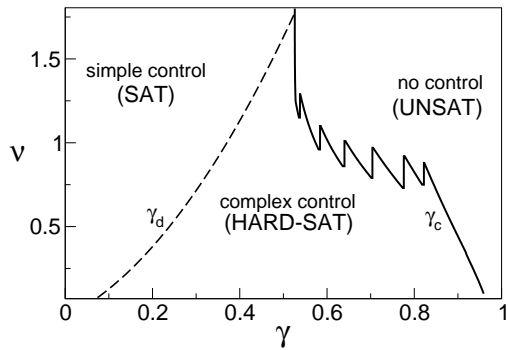}
   \caption{Phase diagram $\gamma - \nu$ for the multi-Poisson
     case. The dashed line, a power law with exponent $\zeta \simeq 1.558$,
     represents the mean value of the numerically evaluated critical parameter
     $\gamma_d (\nu)$ for the simple-complex control transition of network
     realizations with $N=3 \times 10^4$.}
  \label{fig:mp-pd}
\end{figure}
The behavior of GR1 on such a topology is nontrivially different from the
fixed connectivity case.  The main reason for this is that, while
$\Delta_{C}(\gamma)$ is still locally decreasing, many new discontinuities
emerge, due to the influence of clauses with different connectivities. This
gives rise to different phenomena. Firstly, $\Delta_{C}$ can increase globally
with increasing $\gamma$. Indeed, it does increase step-wise with $\gamma$
after $\gamma_{d}$, to decrease again before $\gamma_{c}$. Its discrete jumps
are due to the onset of complex control phases for the different values of
$k$ (Fig.~\ref{fig:mp-delta}).
This fact has an influence on the number of compatible states as a function of
$\gamma$.  Secondly, $\Delta_{C}$ can become negative and then jump back to a
positive value, creating a reentrant UNSAT phase (Fig.~\ref{fig:mp-pd}).
This means that, on average, by increasing the number of constraints one can
pass from unsolvable problems to solvable ones.
%
A heuristic explanation for this counterintuitive fact is that, at fixed $N$,
the addition of a constraint might connect a closed loop in the core to
external free genes, thereby solving a contradiction.  Interestingly, the
simple to complex control boundary is a power-law in $\gamma$
(Fig.~\ref{fig:mp-pd}).

\paragraph{Discussion.}
To conclude, we established a simple framework for the modeling of large scale
transcription networks. It is a compatibility analysis on the constraints
established by transcription.  Its advantage is that, while avoiding to deal
directly with the dynamics, it gives non-trivial results.  In the absence of
an explicit knowledge of the emergent time scales, we feel this is an
appropriate approach, particularly in the Boolean approximation treated here.
From a technical standpoint, GR1 is different from other problems of the
satisfiability kind because of the particular structure of its constraints.
This makes it possible to apply the leaf removal technique, which is
ineffective for other models, such as random-k-Sat~\cite{MPZ02}.
From a general standpoint, our model shows that the ``biological'' complexity
is not simply measured by the number of genes. A more proper indicator is
$\Delta_{C}$ which depends on the order parameter $\nu$, or - roughly - on
the number of transcription factors per gene.  At fixed number of genes, it is
known that this quantity increases in bacteria that need to react to more
environments~\cite{CdL03}.
Imagining that prokaryotes are naturally found in a simple control phase, our
phase diagram predicts an intrinsic limit to this process, represented by the
phase boundary with the complex control phase.
The multi-Poisson case gives an interesting prediction for the behaviour of
this boundary at fixed $N$.  Namely, at criticality, the average number of
constraints scales as a power-law with $\gamma$. This feature, together with
the existence of a core and the predictions on the control exerted by free
genes, can possibly be tested experimentally.

The approach presented here is new, and largely unexplored. It is naturally
fit to study networks with non-Boolean variables and probabilistic
constraints.
It can be of use for models that describe other regulation mechanisms than
just transcription.
More far-reached extensions include evolutionary models.
%
It is not clear yet exactly how useful it can be for the study of concrete
biological networks.  Together with the general trends, a biologically
significant model has to be able to deal with the details of an individual
realization the system.  This is, we think, the main challenge to our
approach, and the direction we are currently exploring.
%
%
%
\begin{acknowledgments}
  We would like to acknowledge interesting discussions with J.~Berg,
  M.~Caselle, L.~Finzi, M.~Leone, A.~Sportiello, P.R.~Tenwolde, R.~Zecchina.
  We thank an anonymous referee for help improving our manuscript.
\end{acknowledgments}

\appendix

\section*{Appendix}

\paragraph{A. GR1 as a Satisfiability problem}
In this appendix we show how a realization of GR1 can be formulated as a
satisfiability problem (Sat)~\cite{M02}, an optimization problem where $N$
Boolean variables are constrained by $O$ conjunctive normal form (CNF)
constraints (i.e. by a Boolean polynomial constructed as a product ($\wedge$)
of $O$ disjunctive monomials ($\vee$)).  Equation (\ref{eq:fp}) is equivalent
to the XOR Boolean constraint
\begin{math}
  I_b = \neg(x_{i(b,0)} \dot{\vee} f_b)
\end{math}.
This can be recast in CNF, as $2^{K_{b}}$ clauses involving $k_{b}$ variables.
Each clause corresponds with a simple map to each line $x_{i(b,1)}, ..,
x_{i(b,K_b)}, x_{i(b,0)}$ (including the output) in the truth table of
$f_{b}$. Namely, if the value of variable $x_{i(b,j)}$ is $1$ in the truth
table line, it will be negated in the CNF clause.  Viceversa, it will be
affirmed if its value is $0$.  The formula
\begin{math}
  I = \bigwedge_{b=1..M} I_b
\end{math},
defines a Satisfiability problem on the variables $x_1,.., x_N$, with
$O=\sum_{b=1..M}2^{K_{b}}$. A realization of GR1 differs from a Sat problem
for the intrinsically asymmetric form of the constraints, which ``force'' the
value of $x_{i(b,0)}$.  Moreover, considering random instances of the problem,
the space of allowed functions of GR1 is much smaller than the corresponding
Sat problem.  For example, for a clause with fixed connectivity $k$, while
there are $2^{2^k}$ possible Boolean functions, all of which can appear in
Sat, only $2^{2^K}$ of these can appear in GR1.
The two above features
make the leaf removal technique useful for the latter model.

\paragraph{B. Average Number of Solutions}
In this appendix, we discuss the average of $\mathcal{N}$ on the realizations
of the constraints $ \{ \vec{I}, \vec{f} \}$, for $K$-GR1. One can write
\begin{displaymath}
  \mathcal{N}(\vec{I}, \vec{f}) = \sum_{\vec{x}} \prod_{b=1}^M
  \delta(x_{i(b,0)}; f_{b}(x_{i(b,1)}, .., x_{i(b,K_b)})).   
\end{displaymath}
Here, the randomness is contained: (i) in the specification of the network
structure, $\vec{I} = (I_1,...,I_M)$, i.e. in the coordinates $i(b,l)$; (ii)
in the specification of the functions $\vec{f}= (f_1,...,f_M) $ within the
class $\mathcal{F}$ of all Boolean functions. An overbar ($\bar{\ ~}$)
indicates an average on both distributions, $p(\vec{I})$ and $p(\vec{f})$.
Taking for $\mathcal{F}$ the class of all Boolean functions, it is
straightforward to obtain
\begin{math}
  \overline{\mathcal{N}} = 2^{N-M} 
\end{math},
independently from the specification of the network structure.
This result remains true considering a sub-family $\mathcal{F}_{\rho}$ of
functions that satisfy
\begin{math}
  \frac{1}{2^{2^K}} \sum_{f \in \mathcal{F}_{\rho} } p(\vec{f}) f(\vec{x}) =
  \rho
\end{math}.
The reason for this is that one finds 
  \begin{math}
    \overline{ \mathcal{N}} = \sum_{\vec{x},\vec{I}} p(\vec{I}) 
    \prod_{b=1}^M \left(
      \rho \delta_{1; x_{i(b,0)}} + (1- \rho) \delta_{0; x_{i(b,0)}} 
    \right) 
  \end{math}.


\begin{thebibliography}{20}
\expandafter\ifx\csname natexlab\endcsname\relax\def\natexlab#1{#1}\fi
\expandafter\ifx\csname bibnamefont\endcsname\relax
  \def\bibnamefont#1{#1}\fi
\expandafter\ifx\csname bibfnamefont\endcsname\relax
  \def\bibfnamefont#1{#1}\fi
\expandafter\ifx\csname citenamefont\endcsname\relax
  \def\citenamefont#1{#1}\fi
\expandafter\ifx\csname url\endcsname\relax
  \def\url#1{\texttt{#1}}\fi
\expandafter\ifx\csname urlprefix\endcsname\relax\def\urlprefix{URL }\fi
\providecommand{\bibinfo}[2]{#2}
\providecommand{\eprint}[2][]{\url{#2}}

\bibitem[{\citenamefont{Babu et~al.}(2004)\citenamefont{Babu, Luscombe,
  Aravind, Gerstein, and Teichmann}}]{BLA+04}
\bibinfo{author}{\bibfnamefont{M.}~\bibnamefont{Babu}},
\emph{et al.}, 
\bibinfo{journal}{Curr Opin Struct Biol}
\textbf{\bibinfo{volume}{14}}, \bibinfo{pages}{283} (\bibinfo{year}{2004}).

\bibitem[{\citenamefont{Herrgard et~al.}(2004)\citenamefont{Herrgard, Covert,
  and Palsson}}]{HCP04}
  \bibinfo{author}{\bibfnamefont{M.}~\bibnamefont{Herrgard}},
\emph{et al.},
  \bibinfo{journal}{Curr Opin Biotechnol} \textbf{\bibinfo{volume}{15}},
  \bibinfo{pages}{70} (\bibinfo{year}{2004}).

\bibitem[{\citenamefont{Ptashne}(1992)}]{Pta92}
\bibinfo{author}{\bibfnamefont{M.}~\bibnamefont{Ptashne}},
  \emph{\bibinfo{title}{A {G}enetic {S}witch}} (\bibinfo{publisher}{Cell Press,
  MA}, \bibinfo{year}{1992}).

\bibitem[{\citenamefont{Buchler et~al.}(2003)\citenamefont{Buchler, Gerland,
  and Hwa}}]{BGH03}
\bibinfo{author}{\bibfnamefont{N.}~\bibnamefont{Buchler}},
\emph{et al.},
\bibinfo{journal}{Proc Natl Acad Sci USA} \textbf{\bibinfo{volume}{100}},
\bibinfo{pages}{5136} (\bibinfo{year}{2003}).

\bibitem[{\citenamefont{Shen-Orr et~al.}(2002)\citenamefont{Shen-Orr, Milo,
  Mangan, and Alon}}]{SMM+02}
\bibinfo{author}{\bibfnamefont{S.}~\bibnamefont{Shen-Orr}},
\emph{et al.},
\bibinfo{journal}{Nat Genet} \textbf{\bibinfo{volume}{31}},
\bibinfo{pages}{64} (\bibinfo{year}{2002}).

\bibitem[{\citenamefont{McAdams and Arkin}(1997)}]{MA97}
\bibinfo{author}{\bibfnamefont{H.}~\bibnamefont{McAdams}} \bibnamefont{and}
  \bibinfo{author}{\bibfnamefont{A.}~\bibnamefont{Arkin}},
  \bibinfo{journal}{Proc Natl Acad Sci USA} \textbf{\bibinfo{volume}{94}},
  \bibinfo{pages}{814} (\bibinfo{year}{1997}).


\bibitem[{\citenamefont{Gillespie}(1977)}]{GD77}
\bibinfo{author}{\bibfnamefont{D.}~\bibnamefont{Gillespie}},
  \bibinfo{journal}{J. Phys. Chem.} \textbf{\bibinfo{volume}{81}},
  \bibinfo{pages}{2340 61} (\bibinfo{year}{1977}).

\bibitem[{\citenamefont{Kauffman}(1993)}]{Kau93}
\bibinfo{author}{\bibfnamefont{S.}~\bibnamefont{Kauffman}},
  \emph{\bibinfo{title}{The {O}rigins of {O}rder}} (\bibinfo{publisher}{Oxford
  Univ. Press, New York}, \bibinfo{year}{1993}).

\bibitem[{\citenamefont{Gershenson}(2004)}]{Ger04}
\bibinfo{author}{\bibfnamefont{C.}~\bibnamefont{Gershenson}}, in
  \emph{\bibinfo{booktitle}{Artificial {L}ife {IX} {W}orkshops and
  {T}utorials}} (\bibinfo{year}{2004}).


\bibitem[{\citenamefont{Mertens}(2002)}]{M02}
\bibinfo{author}{\bibfnamefont{S.}~\bibnamefont{Mertens}},
  \bibinfo{journal}{Comput Sci and Eng}
  \textbf{\bibinfo{volume}{4}}, \bibinfo{pages}{31} (\bibinfo{year}{2002}).

\bibitem[{\citenamefont{Mezard et~al.}(2002)\citenamefont{Mezard, Parisi, and
  Zecchina}}]{MPZ02}
\bibinfo{author}{\bibfnamefont{M.}~\bibnamefont{Mezard}},
\emph{et al.},
  \bibinfo{journal}{Science} \textbf{\bibinfo{volume}{297}},
  \bibinfo{pages}{812} (\bibinfo{year}{2002}).

\bibitem[{\citenamefont{Mezard et~al.}(2003)\citenamefont{Mezard,
  Ricci-Tersenghi, and Zecchina}}]{MRZ03}
\bibinfo{author}{\bibfnamefont{M.}~\bibnamefont{Mezard}},
\emph{et al.}, \bibinfo{journal}{J Stat Phys} \textbf{\bibinfo{volume}{505}}
(\bibinfo{year}{2003}). See also  S.~Cocco \emph{et~al.},
 \bibinfo{journal}{Phys Rev Lett} \textbf{\bibinfo{volume}{90}}, 047205
  (\bibinfo{year}{2003}) and  L.~Correale \emph{et~al.},
  cond-mat/0412443.

\bibitem[{\citenamefont{Shea and Ackers}(1985)}]{SA85}
\bibinfo{author}{\bibfnamefont{M.}~\bibnamefont{Shea}} \bibnamefont{and}
  \bibinfo{author}{\bibfnamefont{G.}~\bibnamefont{Ackers}}, \bibinfo{journal}{J
  Mol Biol} \textbf{\bibinfo{volume}{181}}, \bibinfo{pages}{211}
  (\bibinfo{year}{1985}).

\bibitem[{\citenamefont{Cosentino~Lagomarsino
  et~al.}(2005)\citenamefont{Cosentino~Lagomarsino, Bassetti, and
  Jona}}]{CLB05}
\bibinfo{author}{\bibfnamefont{M.}~\bibnamefont{Cosentino~Lagomarsino}},
\emph{et al.}, (\bibinfo{year}{2005}), \bibinfo{note}{q-bio.MN/0502017}.


\bibitem{nota_sav} We should note that the conventional average of
  $\mathcal{N}$ might be biased by the weight of exceptions~\cite{MPV87}.  To
  access the typical behavior of the system, the correct quantity to compute
  is the ``quenched average'', $\overline{\log{\mathcal{N}}}$, usually
  accessed with the replica, or similar methods~\cite{MPZ02}.  In the case
  under exam we have computed the \emph{annealed} average
  $\log{\overline{\mathcal{N}}}$ (in general $\overline{\log{\mathcal{N}}}
  \leq \log{\overline{\mathcal{N}}}$).  For K-GR1 with Poisson distributed
  out-degree, we have shown the presence of a self-averaging property for
  $\mathcal{N}$. 
That is, the quantity
\begin{math}
  (\overline{[\mathcal{N}]^{2}} - \left[\overline{\mathcal{N}}\right]^2) /
  (\left[\overline{\mathcal{N}}\right]^2)  
\end{math}
vanishes in the thermodynamic limit $N\to\infty$ so that, in the simple and
complex control regimes, an equality holds between quenched and annealed
average~\cite{CLB05}.
  

\bibitem[{\citenamefont{Mezard et~al.}(1987)\citenamefont{Mezard, Parisi, and
      Virasoro}}]{MPV87}
  \bibinfo{author}{\bibfnamefont{M.}~\bibnamefont{Mezard}}, \emph{et al.},
  \emph{\bibinfo{title}{Spin {G}lass {T}heory and {B}eyond}}
  (\bibinfo{publisher}{World Scientific, Singapore}, \bibinfo{year}{1987}).
  

\bibitem[{\citenamefont{Cases et~al.}(2003)\citenamefont{Cases, de~Lorenzo, and
  Ouzounis}}]{CdL03}
\bibinfo{author}{\bibfnamefont{I.}~\bibnamefont{Cases}},
\emph{et al.}, 
\bibinfo{journal}{Trends Microbiol}
\textbf{\bibinfo{volume}{11}}, \bibinfo{pages}{248} (\bibinfo{year}{2003}).

\bibitem[{\citenamefont{van Nimwegen}(2003)}]{vN03}
\bibinfo{author}{\bibfnamefont{E.}~\bibnamefont{van Nimwegen}},
  \bibinfo{journal}{Trends Genet} \textbf{\bibinfo{volume}{19}},
  \bibinfo{pages}{479} (\bibinfo{year}{2003}).


\end{thebibliography}

\end{document}